\documentclass[conference]{IEEEtran}
\usepackage[utf8]{inputenc}
\usepackage{xcolor,amsmath,amsfonts,amssymb,graphicx}
\usepackage{algorithm}
\usepackage{algpseudocode}%
\usepackage{amsthm}
\newtheorem{theorem}{Theorem}
\newtheorem{lemma}{Lemma}

\theoremstyle{remark}
\newtheorem{rem}{Remark}
\title{INTERPOL: Information Theoretically Verifiable Polynomial Evaluation}
\author{}
\date{August 2018}
\author{\IEEEauthorblockN{Saeid Sahraei and A. Salman Avestimehr}
\IEEEauthorblockA{\textit{Department of Electrical Engineering, University of Southern California, Los Angeles, CA}}
}
\begin{document}
\maketitle

\begin{abstract}
    We study the problem of verifiable polynomial evaluation in the user-server and multi-party setups. 
    We propose {INTERPOL}, an information-theoretically verifiable algorithm that allows a user to delegate the evaluation of a polynomial to a server, and verify the correctness of the results with high probability and in sublinear complexity.
    Compared to the existing approaches which typically rely on cryptographic assumptions, {INTERPOL} stands out in that it does not assume any computational limitation on the server. 
    {INTERPOL} relies on decomposition of polynomial evaluation into two matrix multiplications, and injection of computation redundancy in the form of locally computed parities with secret coefficients for verification.
We show that {INTERPOL} has several desirable properties such as adaptivity and public verifiability.
Furthermore, by generalizing {INTERPOL} to a multi-party setting consisting of a network of $n$ untrusted nodes, where each node is interested in evaluating the same polynomial, we demonstrate that we can achieve an overall computational complexity comparable to a trusted setup, while guaranteeing information-theoretic verification at each node.  
\end{abstract}
\section{Introduction} 
Cloud and edge computing are rapidly growing in popularity by enabling users to simply offload their compute-intensive tasks via the Internet. But, how can we make sure that the computations are done correctly in the cloud? For example, there may be dishonest servers in the cloud that may return plausible (and potentially misleading) results without performing the actual work. This critical problem has motivated the formalization of  Verifiable Computation, which is to enable offloading of computations to untrusted servers while maintaining verifiable results (e.g., \cite{gennaro2010non,benabbas2011verifiable}).

\begin{figure}
\centering
\includegraphics[scale=0.108]{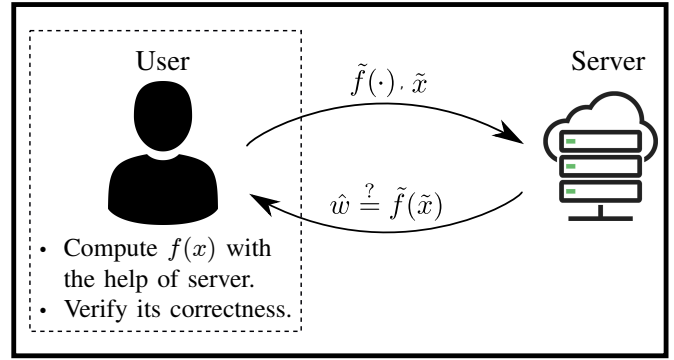}
\caption{Illustration of verifiable computation, in which a user wishes to offload the computation of $f(x)$ to a server, and efficiently verify the correctness of the results.}
\label{fig:toy}
\end{figure}

More specifically, as depicted in Figure~\ref{fig:toy}, in the problem of verifiable computing a user is interested in evaluating $f(x)$. After some preprocessing, he will reveal a function $\tilde{f}$ and an input value $\tilde{x}$ to the server. The server will be in charge of returning $m = \tilde{f}({\tilde{x}})$. Given a returned value $\hat{m}$, the user must be able to (i) verify that indeed $\hat{m} = \tilde{f}({\tilde{x}})$, and (ii) recover the value of $f(x)$. He must be able to perform both of these tasks in substantially smaller complexity than the original computation of $f(x)$.

The problem of verifiable computation has a rich history in the literature. A large body of work focuses on computation of arbitrary functions, by relying on Interactive Proofs \cite{babai1985trading}, Probabilistically Checkable Proofs \cite{kilian1992note,goldwasser2008delegating}, Fully Homomorphic Encryption \cite{gennaro2010non}, or converting arithmetic circuits into Quadratic Arithmetic Programs \cite{parno2013pinocchio,gennaro2013quadratic}. Despite their theoretical beauty, many of these works are still far from being practically implementable. As a result, there has been several recent efforts to address the computation of specific functions which are of popular demand, such as polynomial evaluation and matrix multiplication \cite{elkhiyaoui2016efficient,fiore2012publicly,benabbas2011verifiable}. The focus of this line of work is to present algorithms which can be implemented and provide satisfactory performance guarantees, at the expense of generality. 

Our main contribution in this paper is the development of a new algorithm, named \textsf{INTERPOL}\footnote{\textsf{INTERPOL} stands for \textbf{in}formation \textbf{t}heoretically v\textbf{er}ifiable \textbf{pol}ynomial evaluation}, for verifiable polynomial computing. The distinguishing feature of  \textsf{INTERPOL}  is that it does not rely on any cryptographic assumption. As a result, even a computationally unbounded adversarial server, or one equipped with a quantum computer cannot compromise the security of the system. To the best of our knowledge, \textsf{INTERPOL} is the first information-theoretically secure algorithm for verifiable polynomial computation.

 To accomplish this, we first transform the problem of polynomial evaluation to two matrix multiplications, the first performed by the server and the second by the user. The complexity of the first matrix multiplication is linear in $k$, where $k$ is the degree of the polynomial that needs to be evaluated. Hence, the complexity of the server remains the same as that of a polynomial computation. On the other hand, the complexity of the second matrix multiplication, which is done by the user, is only $O(\sqrt{k})$. Furthermore, we provide a simple mechanism for the user to verify the correctness of the first matrix multiplication by performing several parity checks with secret coefficients. This verification too can be done in  $O(\sqrt{k})$. As a result, the overall complexity of the user will be $O(\sqrt{k})$, which is much smaller than evaluating the polynomial. The security of  \textsf{INTERPOL}  merely relies on hiding the secret coefficients of the parity checks from the server, and thus cannot be compromised by a computationally unbounded server.
 

The second feature of  \textsf{INTERPOL}  is that it is publicly verifiable \cite{zhang2017new}, in the sense that not only the user, but any other node in the network can perform the verification and decide whether the result of computation is correct. This property allows us to extend our results to a network of $n$ nodes where all the nodes are interested in evaluating the same polynomial at a given input. This setup is indeed very common, for instance in the context of blockchain, where all the full nodes in the network examine a newly mined block to check the validity of the transactions included therein. This process can be formulated as a polynomial evaluation \cite{polyshard} which can be captured by our model. Our approach to this problem is as follows. We require each node to perform a small part of the computation. The nodes then exchange these intermediate results and verify the correctness of the results provided by other nodes in the network. Following this procedure, and assuming that the number of nodes in the network is substantially smaller than the dimensionality of the problem, we can reduce the overall computational complexity of the network by a factor of $n$, compared to the scenario where each node performs the computation individually.

\section{Problem Statement}
\label{sec:statement}
A user wishes to delegate the computation of a polynomial ${f}(x)$ of degree $k$ over some finite field $\mathbb{F}_q$, at a series of input values $x\in \{x_1,\dots ,x_{max}\}$ to a server. We assume that both $f$ and the set  $\{x_1,x_2,\dots,x_{max}\}$ are known globally. The user must be able to verify, with high probability and in low complexity, the correctness of the result provided by the server. We consider an amortized model \cite{gennaro2010non} where the user can afford to perform a one-time computionally heavy task. A commonly-adapted assumption in the literature is that the number of input values at which we are interested in evaluating $f(x)$ is so large that this initialization cost becomes negligible {\it per round}.

Let us represent the overall complexity of the user for one round of computation by ${\cal C}$, and let ${\cal P}$ be the probability of error of the user, i.e., the probability that he would accept a false result as valid. We are interested in characterizing the tradeoff between ${\cal P}$ and ${\cal C}$. In particular, our goal is to achieve a ${\cal C}$ which is sublinear in the degree of the polynomial, and a probability of error that vanishes as the field size grows large. We will now make these definitions precise.

\begin{figure}
    \centering
    \includegraphics[scale = 0.18]{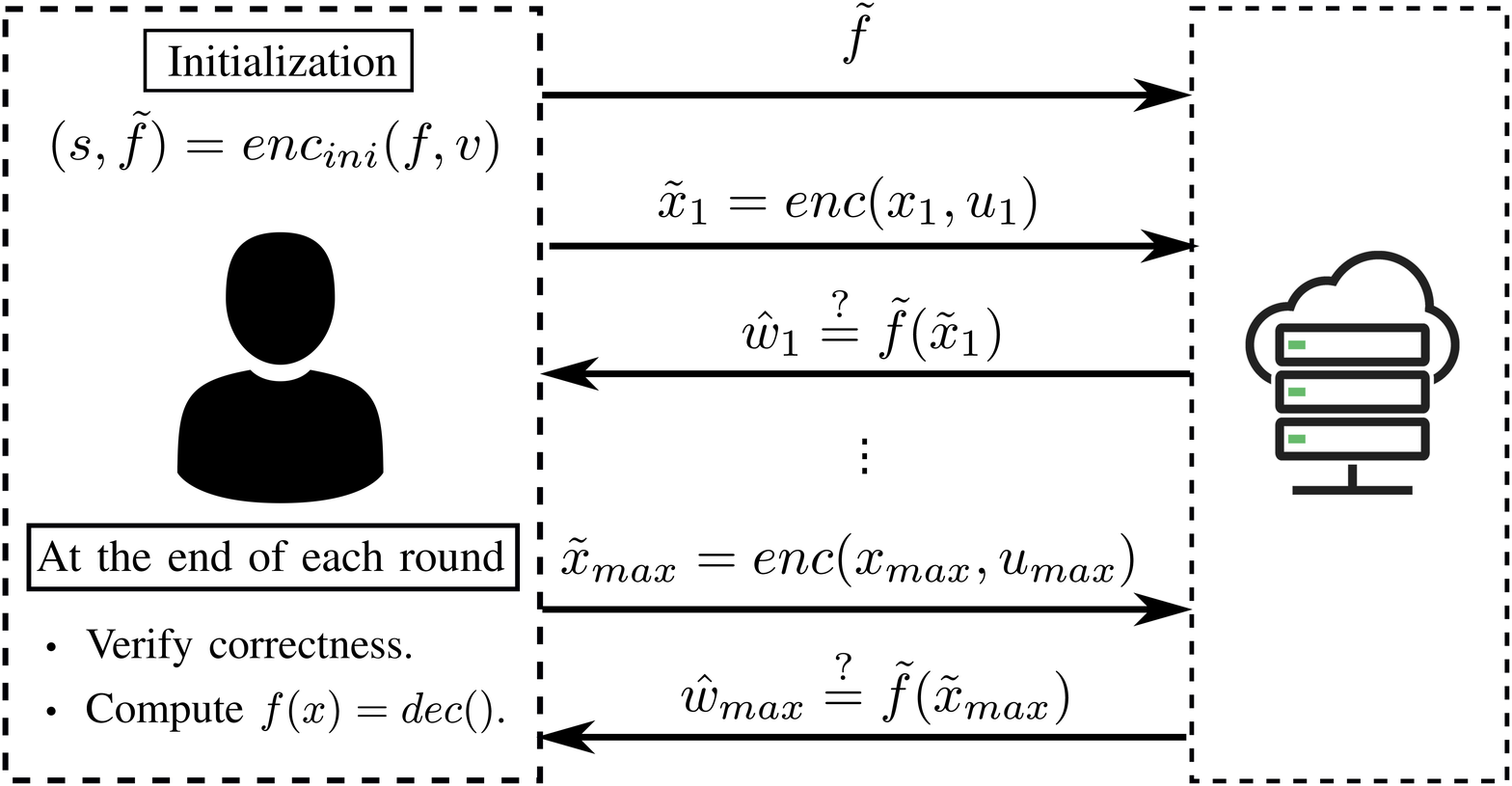}
    \caption{Illustration of various stages of verifiable computing. {\bf Initialization:} the user generates a random variable $v$, and computes $(s,\tilde{f}) = enc_{ini}(f,v)$. He keeps $s$ private, but reveals $\tilde{f}$ to the server. {\bf At round $i$}: the user generates a random variable $u_i$ and computes $\tilde{x}_i = enc(x_i,u_i)$. He reveals $\tilde{x}_i$ to the server and demands the computation of $\tilde{f}(\tilde{x}_i)$. Once he receives the response from the server, he verifies the correctness of the result, and subsequently recovers $f(x)$.}
    \label{fig:user_server}
\end{figure}
\subsubsection*{{\bf The initialization phase}} The user generates a random variable $v$ according to some distribution $P_{V}$ and computes $(s,\tilde{f}) = enc_{ini}(f,v)$. He keeps $s$ private but reveals $\tilde{f}$ to the server.
\subsubsection*{{\bf At round $i$}} The user wishes to recover $f(x_i)$. The computation is done in three steps. 
    \begin{itemize}
        \item The user computes $\tilde{x}_i = enc(x_i,u_i)$ for some random variable $u_i$ independent of $v$ and $u_{[i-1]}$, with marginal distribution $P_{ U}$. He reveals $\tilde{x}_i$ to the server and requests the computation of $w = \tilde{f}(\tilde{x}_i)$.
        \item The server then returns a possibly randomized function \begin{eqnarray}
        \hat{w} = comp(\tilde{f},f,\tilde{x}_{[i]},{x}_{[max]}).\end{eqnarray} 
        \item The user computes two functions. First, a verification bit $b = ver(\hat{w},s,f,x_i)$ is computed. If $b = 0$, the user rejects the result of the computation. Otherwise, he will aim at recovering the evaluation of $f$ at $x$ by computing $\hat{r} = dec(\hat{w},s,f,x_i)$. He will accept $\hat{r} = r = f(x_i)$. 
    \end{itemize} 


The algorithm described above must satisfy  the following properties. 
\begin{itemize}
\item {\bf Correctness:} If the server is honest, and $\hat{w} = w$, then the verification process must pass and the user must be able to recover $f(x)$. In other words, for any $x\in \{x_1,x_2,\dots,x_{max}\}$ we must have
\begin{eqnarray}
 dec({w}, s,f,x) &=& f(x),\\
ver({w}, s, f,x) &=& 1.
\end{eqnarray}
\item {\bf (Information-Theoretic) Soundness:} If the server is dishonest, the verification process must fail with high probability. More formally, suppose at round $i$ the server returns $\hat{w} \neq \tilde{f}(\tilde{x})$ where $\hat{w}$ could only depend on $\tilde{x}_{[i]}, x_{[max]}$, $\tilde{f}$ and $f $. Then, with high probability we must have $ver(\hat{w}, s, f,x_i) = 0$. In other words,
${\cal P} = o(1)$ where
\small
\begin{eqnarray*}
{\cal P}\stackrel{\bigtriangleup }{=}\max_{f,x_{[max]},comp(\cdot)}{\mathbb P}(ver(\hat{w}, s, f,x_i) &=& 1 |\hat{w} \neq w).
\end{eqnarray*}
\normalsize
The term $o(1)$ must vanish as the size of the field $q$ grows large. Note that we are considering a worst case scenario over all possible functions $comp(\cdot)$. This implies that the soundness property must be {\it information-theoretic}: it must hold for any server, regardless of his computational budget.
\item {\bf Efficient Verification and Recovery:}
The entire process of encoding the input, recovering the value of $f(x)$ from $w$ and verifying the result of the computation must be substantially easier than performing the original polynomial evaluation. More formally, define $c_{enc}$, $c_{ver}$ and $c_{dec}$ as the maximum complexity of computing $enc(x,u)$, $ver({w}, s, f,x)$ and $dec({w}, s, f,x)$ respectively. This maximum is taken over all polynomials of degree $k$ over $\mathbb{F}_q$ and all possible set of input values. 
We must have ${\cal C} = o(k)$ where
\begin{eqnarray}
{\cal C} \stackrel{\bigtriangleup}{=}c_{enc}+c_{ver} + c_{dec}.
\end{eqnarray}
\item {\bf Efficient Computation:} Even though we do not assume any limitation on the computation power of the server, we require the complexity of computing $\tilde{f}(\tilde{x})$ to be comparable to the complexity of computing $f(x)$. More precisely, we require that 
$c_{\tilde{{f}}} = \tilde{O}(k)$, where $\tilde{O}$ can hide a polylogarithmic term. 
Imposing this (or a slightly looser) restriction is crucial in designing practical algorithms.
\end{itemize}
Given the requirements above, we ask what is the tradeoff between the parameters ${\cal P}$ and ${\cal C}$ among all possible  strategies that achieve $c_{\tilde{f}} = \tilde{O}(k)$.

\begin{figure}%
\centering
    \includegraphics[scale= 0.22]{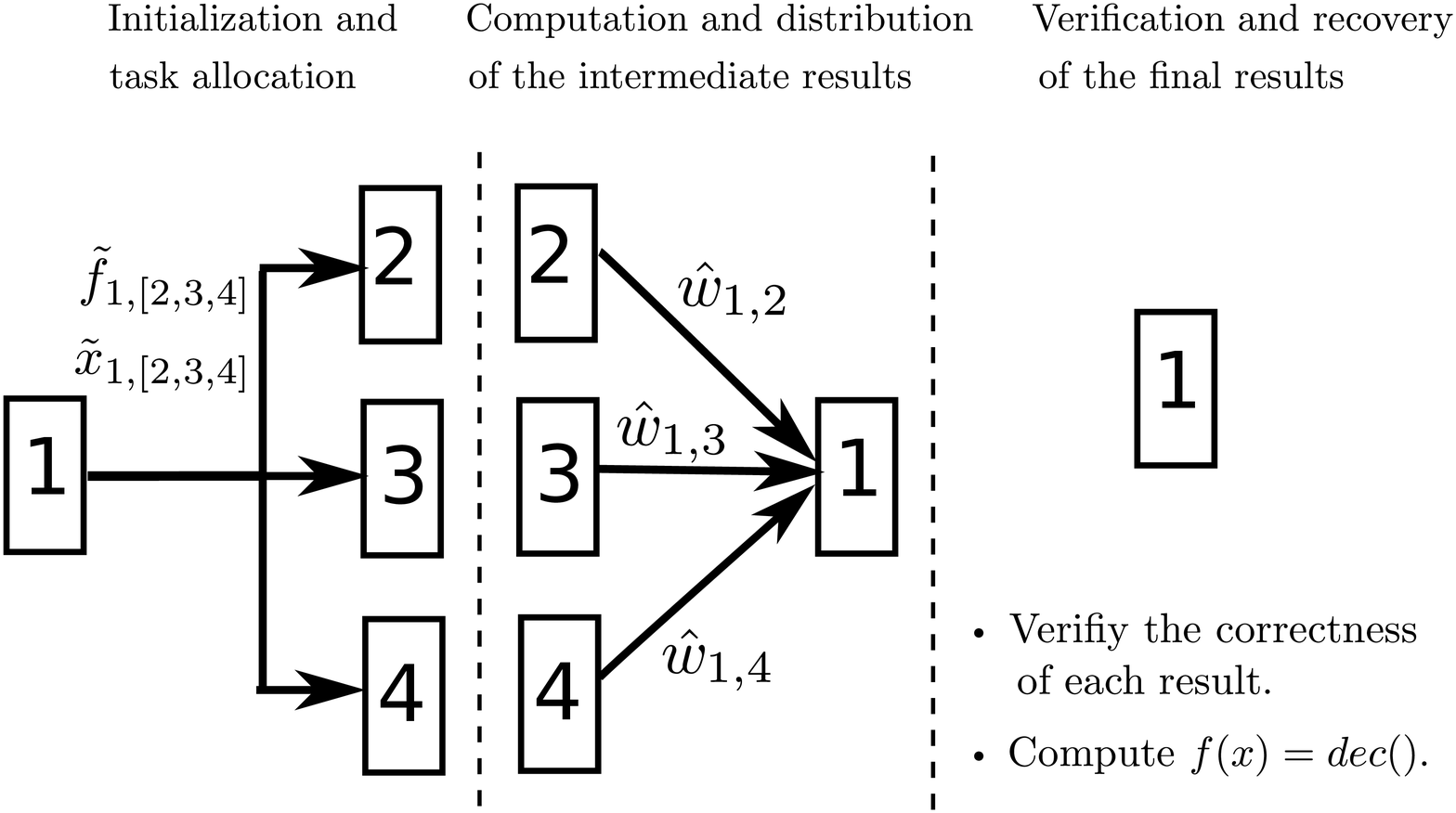}
    \caption{In multi-party verifiable computation, each pair of nodes can be viewed as two user-server instances. For each instance, one node performs the computation and the other verifies the result.}
    \label{fig:multi_party}
\end{figure}
\subsection{Multi-party setup}
\label{sec:def_multi}
As a secondary model, we consider a network of $n$ nodes which are all interested in computing $f(x)$ over a series of input values. We assume that $f(\cdot)$ and the set of inputs are publicly known. Furthermore, $k\gg n$, that is, the dimensionality of the problem is much larger than the number of nodes in the network. In its general form, we can decompose the network into $n(n-1)$ user-server pairs $\{(\ell,j),\ell \in [n], j\in[n], \ell\neq j\}$. For each pair $(\ell,j)$, node $j$ will be in charge of computing a specific task and server $\ell$ will verify the correctness of the result. 
 \subsubsection*{\bf Initialization} Each node $\ell$ generates a random variable $v_\ell$ based on some distribution $P_{{\cal V}_\ell}$ and computes $(s_\ell,\tilde{f}_{\ell,[n]\backslash\{\ell\}}) = enc_{\ell,ini}(f,v)$. If node $\ell$ is honest, we assume that $v_\ell$ is independent of $(v_{[n]\backslash\{\ell\}})$. Dishonest nodes may collude and choose their random variables based on some joint distribution. Each honest node $\ell$ keeps $s_\ell$ private. Each node $\ell$ reveal $\tilde{f}_{\ell,[n]\backslash\{\ell\}}$ to everyone.

\subsubsection*{\bf At round $i$} Each node $\ell$  wishes to recover $f(x_i)$. 
    \begin{itemize}
       \item Node $\ell$ computes $\tilde{x}_{i,\ell,[n]\backslash\{\ell\}} = enc_{\ell}(x_i,u_{i,\ell})$ for some random variables $u_{i,\ell}$. He then broadcasts $\tilde{x}_{i,\ell,[n]\backslash\{\ell\}}$ to all the nodes in the network. Node $j$ will be in charge of computing $w_{\ell,j} = \tilde{f}_{\ell,j}(\tilde{x}_{i,\ell,j})$ for all $\ell \in [n]\backslash \{j\}$.
       \item Node $j$ will computes a possibly randomized function $$\hat{w}_{[n]\backslash\{j\},j} = comp(\tilde{f}_{[n],[n]},f,\tilde{x}_{[i],[n],[n]},x_{[max]},v_j,u_{[i],j})$$ and broadcasts the result of computation to every node in the network. 
        \item Each node $\ell$ computes two functions. First, a verification string of bits $b_{\ell,j} = ver_{\ell,j}(\hat{w}_{\ell,[n]\backslash\{\ell\}},s_\ell,f,x_i)$ for $j\in[n]\backslash\{\ell\}$ is computed. If there exists some $j\in [n]\backslash\{\ell\}$ such that $b_{\ell,j}=0$, node $\ell$ will reject the result of computation provided by node $j$. Otherwise, he will aim at recovering the evaluation of $f$ at $x$ by computing $\hat{r} = dec_\ell(\hat{w}_{\ell,[n]\backslash\{\ell\}},s_\ell,f,x_i)$. He will accept $\hat{r} = r = f(x)$.
   \end{itemize}
   
The algorithm must satisfy the correctness and soundness properties defined similarly to the user-server setup.
\begin{itemize}
    \item {\bf Correctness:} \begin{eqnarray}
 dec_\ell({w}_{\ell,[n]\backslash\{\ell\}}, s_\ell, f,x) &=& f(x),\;\; \forall \ell\\
ver_\ell({w}_{\ell,[n]\backslash\{\ell\}}, s_\ell, f,x) &=& 1, \;\;\forall \ell.
\end{eqnarray}
\item {\bf (Information-Theoretic) Soundness:} ${\cal P} =o(1)$ where 

\begin{eqnarray*}
{\cal P}_{\ell,j} \stackrel{\bigtriangleup}{=} \max_{f,x_{[max]},comp(\cdot)}{\mathbb P}(ver_{\ell,j}(\hat{w}_{\ell,[n]\backslash\{\ell\}},s_\ell,f, x_i) \\=1| \hat{w}_{\ell,j} \neq w_{\ell,j})
\end{eqnarray*}
and ${\cal P} = \max_{\ell,j} {\cal P}_{\ell,j}$.
In particular, this property must hold regardless of the number of malicious nodes in the network and their state of collusion.
\end{itemize}

For this model, we are interested in characterizing the trade-off between ${\cal P}$ and the worst-case complexity of the nodes,
\begin{eqnarray*}{\cal C} = \max_{\ell}[c_{enc_\ell} + c_{\tilde{f}_{[n]\backslash\{\ell\},\ell}} + \sum_{j}c_{ver_{\ell,j}} + c_{dec_\ell}].
\end{eqnarray*}

\section{Main Results}
\label{sec:main}
The first contribution of this work is to propose  \textsf{INTERPOL}, an algorithm that allows a user to recover the evaluation of a polynomial $f(x) = a_0 + a_1x + \dots + a_{k-1}x^{k-1}$ in complexity $O(\sqrt{k})$ with the help of a server. Furthermore, the user can verify the correctness of the computation provided by the server in $O(c\sqrt{k})$ for some constant $c$. Essential to the correctness of \textsf{INTERPOL}  is the widely adapted assumption of amortized cost: the user is allowed to perform a one-time heavy computation knowing that this cost will break down over the evaluations of $f(x)$ at $x\in \{x_1,x_2,\dots,x_{max}\}$, and can be neglected \cite{gennaro2010non}. 
\begin{theorem}
There exists an algorithm which allows a user to compute a polynomial $f(\cdot)$ of degree $k$ over $\mathbb{F}_q$ at an input $x$ with the help of a server, which has the following properties.
\begin{itemize}
    \item The user can verify the correctness of the results provided by the server with an error probability of $\frac{1}{q^c}$, where $c$ is an arbitrary constant chosen by the user.
    \item  The computation complexities of the user and the server are $O(c\sqrt{k})$ and $O(k)$ respectively.
\end{itemize}
\label{thm:interpol}
\end{theorem}
\begin{rem}
To prove Theorem \ref{thm:interpol}, we propose \textsf{INTERPOL}, presented in Section \ref{sec:interpol}, which relies on decomposition of polynomial evaluation into two matrix multiplications and injection of computation redundancy in the form of locally computed parities with secret coefficients.
\end{rem}
\begin{rem}
The verification process used in \textsf{INTERPOL} is {\it information-theoretic} as opposed to the cryptographic approach that is commonly adapted in the literature. This means that even a computationally unbounded adversarial server will not be able to compromise the security of the system.
\end{rem}
\begin{rem}Another important property of \textsf{INTERPOL} is its {\it adaptivity} \cite{benabbas2011verifiable}. This refers to the fact that the algorithm remains secure even when the adversarial server knows whether the user has accepted or rejected his responses to the previous queries. We will see in Section \ref{sec:adaptivity} that the probability that a server who receives such feedback can bypass the verification test at least once in $m$ rounds of computation is $\frac{m}{q^c}$. Note that this is only a marginal increase compared to an elementary server who receives no feedback at all, and who chooses to return $m$ random outputs in response to the $m$ queries, thereby, achieving a probability of success of $1- (1 - \frac{1}{q^c})^m$.
\end{rem}
\begin{rem} \textsf{INTERPOL} is {\it publicly verifiable}. This means that not only the user but any other node in the network will be able to verify the correctness of the results, without having to trust the user or the server.
\end{rem}
The fact that  \textsf{INTERPOL}  is publicly verifiable enables us to generalize our results to a multi-party setting as our second contribution. We propose an algorithm that allows every node in a network of $n$ nodes to recover, with high confidence, the result of a polynomial $f(x)$.
\begin{theorem}
For the multi-party setup in Section \ref{sec:def_multi}, there exists a distributed algorithm which allows each node in a network of $n$ nodes to compute a polynomial $f(\cdot)$ of degree $k$ over $\mathbb{F}_q$ at an input $x$, with the following properties.
\begin{itemize}
\item Each node can verify the correctness of the final result with an error probability of $\frac{1}{q^c}$, where $c$ is an arbitrary constant.
    \item  The computation complexity of each node is $O(\frac{k}{n} + cn\sqrt{k})$.
\end{itemize}
\label{thm:interpol_multi}
\end{theorem}
\begin{rem}
To prove Theorem \ref{thm:interpol_multi}, we propose a multi-party variation of \textsf{INTERPOL}, presented in Section \ref{sec:interpol_multi}, which divides the task of polynomial evaluation into $n$ parallel tasks, each performed by one user and verified by the remaining users, following a similar approach to the user-server setup. 
\end{rem}
\begin{rem}In the absence of trust, a naive (but common) approach is for each node in the network to individually compute $f(x)$, which implies an overall complexity of $O(kn)$. By contrast, our approach only requires $O(\frac{k}{n})$  computation per node, or $O(k)$ computation overall.\footnote{As discussed in Section \ref{sec:def_multi}, we assume that $n$ can be neglected compared to $k$. As a result, the first term in $O(k/n + cn\sqrt{k})$ is dominant.} Given that $O(k)$ computation is needed for evaluating an arbitrary polynomial of degree $k$, we can observe that our algorithm is order-wise as efficient as in a trusted setup. 
\end{rem}
\begin{rem}
Our multi-party algorithm can be directly applied to a blockchain network, where all the full nodes wish to verify the validity of a newly mined block of transactions. Without loss of generality, one can model this process as a polynomial evaluation  task  \cite{polyshard}. As the length of the distributed ledger increases, this process grows in complexity. Our multi-party algorithm proposes a natural solution to this problem. Instead of individually validating the blocks, nodes in the network {\it divide} the task of polynomial evaluation among themselves. Subsequently, each node carries out a small amount of computation in order to validate the results provided by the other nodes. If it is detected that a node has provided false results, other nodes can simply redo the computation and prohibit the malicious node from participating in the following rounds.
\end{rem}

\subsection{Comparison with Prior Works}
The problem of verifiable computation has a rich history. Here, we suffice to address the works which are more prominent or closely related to our contribution. One of the first non-interactive verifiable computation algorithms was proposed in \cite{gennaro2010non}. The authors observe that Yao's Garbled circuit \cite{yao1982protocols,lindell2009proof} which was originally designed for two-party secure computation, can also perform a one-time verifiable computation of arbitrary functions. In order to make the circuit reusable, the authors encode the inputs to the circuit with Fully Homomorphic Encryption (FHE). But due to this reliance on FHE, this algorithm is of limited practical interest. An alternative approach is proposed in \cite{parno2013pinocchio}, where the authors represent an arbitrary C code as an arithmetic circuit, which is then converted into a Quadratic Arithmetic Program (QAP) \cite{gennaro2013quadratic}. 
In simple words, the server first evaluates the circuit, which provides the required coefficients for the QAP. He will then generate a proof by evaluating the QAP at a secret value, ``in the exponent". With the help of a bilinear map, the user can check the correctness of the results in constant time by verifying that the returned values satisfy a certain identity.

Other works in the literature focus on the verifiable computation of specific functions, such as matrix multiplication \cite{zhang2014efficient}, modular exponentiation \cite{chen2014new} and polynomial evaluation \cite{benabbas2011verifiable, fiore2012publicly,backes2013verifiable,elkhiyaoui2016efficient}. By sacrificing the generality of the algorithm, this line of research aims at designing verifiable computation schemes which are efficient and practical. 

We will now describe in more details some of the algorithms which are specifically designed for polynomial evaluation. 
In \cite{benabbas2011verifiable} the authors present one of the first efficient verifiable polynomial evaluation algorithms. Put simply, the server is provided with two vectors $[a_0,\dots,a_{k-1}]$ and $[g^{ca_0 + r_0},\dots,g^{ca_{k-1} + r_{k-1}}]$ where $g$ is a generator of the field and $[r_0,\dots,r_{k-1}]$ is a pseudorandom sequence which satisfies the {\it closed form efficiency} property. This property implies that the user can compute $r(x) = r_0 + r_1x + \dots + r_{k-1}x^{k-1}$ in sublinear time in $k$. Subsequently, once the server returns both $f(x) = a_0 + a_1x + \dots + a_{k-1}x^{k-1}$, and $h(x) = g^{{ca_0 + r_0} +(ca_1 + r_1)x + \dots + (ca_{k-1} + r_{k-1})x^{k-1}}$ the user can easily check whether $h(x) = g^{cf(x) + r(x)}$. 
The complexity of the user under this assumption is $O(log(k))$. 

The algorithm presented in \cite{benabbas2011verifiable} is only verifiable by the user. In \cite{fiore2012publicly} the authors present a modification of this algorithm which admits public verifiability. In this case, the user provides a public verification key that any third party can use to check the correctness of the computation.

The authors in \cite{elkhiyaoui2016efficient} suggest that the user starts by generating a polynomial $b(x) = x^2 + b_0$ where $b_0$ is uniformly random over the field. The user then divides $f(x)$ by $g(x)$ to find the quotient $q(x) = q_0 + q_1x + \dots + q_{k-3}x^{k-3}$ and the remainder $r(x) = r_0 + r_1x$ such that $f(x) = b(x)q(x) + r(x)$. The server is provided with the two vectors $[a_0,\dots,a_{k-1}]$ and $[g^{q_0},\dots,g^{q_{k-3}}]$ and is asked to compute $f(x)$ and $g^{q(x)}$. Upon receiving these two values, the user can check whether $g^{f(x)} = (g^{q(x)})^{b(x)}g^{r(x)}$ in constant time. 
This algorithm is also publicly verifiable. 

By comparison to the works mentioned above,  \textsf{INTERPOL}  stands out in that it is information-theoretic and relies on no cryptographic assumption. The verification time of  \textsf{INTERPOL}  is $O(c\sqrt{k})$ where $k$ is the degree of the polynomial. Despite being significantly more efficient than performing the original computation,  \textsf{INTERPOL}  has the disadvantage of being slower than the existing cryptographic works which are specifically tailored to polynomial evaluation and typically run in $\log(k)$ or even constant time. 

{In the remaining of the paper, we represent random variables with capital letters and their realizations with lower case letters. We reserve bold font for vectors and regular font for scalars. Finally, we represent all the matrices with capital Greek alphabet.}

\section{Description of \textsf{INTERPOL} and Proof of Theorem \ref{thm:interpol}}
\label{sec:interpol}
In this section we describe  \textsf{INTERPOL} for polynomial evaluation in a user-server setup. Suppose the user wishes to evaluate $f(x) = a_0 + a_1 x + \dots a_{k-1}x^{k-1}$ at some input value $x$. For simplicity assume that $k = s^2$. The algorithm can easily be generalized to the case when $k$ is not a complete square. Consider the vector $[a_0 ,\dots , a_{k-1}]$. Break this vector down into $s$ consecutive chunks of length $s$ each, and rename the elements as follows: $a_{i,j} = a_{i\times s + j}$ for $i\in [0:s-1]$ and $j\in[0:s-1]$. Define the $s\times s$ matrix ${\Delta}$ as $\Delta = [a_{i,j}]$.
We can now write
\begin{eqnarray*}
f(x) = \begin{bmatrix}
1 & x^s &\dots& x^{s(s-1)}  
\end{bmatrix}
\Delta\begin{bmatrix}
1 & x &\dots& x^{s-1}  
\end{bmatrix}^T.
\end{eqnarray*}
The user delegates the computation of ${\bf b}:=\Delta\begin{bmatrix}
1 & x &\dots& x^{s-1}  
\end{bmatrix}^T $ to the server following an algorithm that we will describe shortly. After this phase, the user computes
$f(x) = \begin{bmatrix}
1 & x^s &\dots& x^{s(s-1)}  
\end{bmatrix} {\bf b}.
$ Note that this can be done in $O(s)$ since each element of $\begin{bmatrix}
1 & x^s &\dots& x^{s(s-1)}  
\end{bmatrix}$ can be computed in constant time\footnote{The $j$'th element of the vector can be computed by one multiplication $x^{(j-1)s} = x^{(j-2)s}x^s$, since $x^{(j-2)s}$ and $x^s$ have been previously computed.}, (except $x^s$ which can be computed in $\log(s)$).  If the user can verify and recover the value of ${\bf b}$ in complexity $O(cs)$ for some constant $c$, the overall complexity of the user will be $O(c\sqrt{k})$. We will now show how to accomplish this. 

Consider a general problem of verifiable matrix-vector multiplication. Suppose the user intends to find $\Delta {\bf z}$ where $\Delta$ is a square $s\times s$ matrix and ${\bf z}$ is a column vector.
The user starts by choosing a constant $c>0$. As we will see soon, a larger $c$ implies a higher complexity for the user, but results in an enhanced security. The user generates a random $c \times s$ matrix $\Lambda$ of uniform i.i.d. elements over the field.
The user then performs a one-time computation of $\Gamma = \Lambda \Delta$. 

Note that this requires $O(cs^2)$ computation. This may seem substantial at first, but one should note that this cost will be amortized, since the user does not need to update $\Gamma$ for every input vector ${\bf z}$.

The user keeps $\Lambda$ and $\Gamma$ as secret. He simply reveals $\Delta$ and ${\bf z}$ to the server and asks him to compute ${\bf w} =\Delta {\bf z}$. After receiving $\hat{{\bf w}}$ from the server, the user checks whether $\Lambda \hat{{\bf w}} = \Gamma {\bf z}$. Note that both $\Lambda \hat{{\bf w}}$ and $ \Gamma {\bf z}$ can be computed in $O(cs)$ which is sublinear in the complexity of computing $\Delta {\bf z}$ directly.

If the equality $\Lambda \hat{{\bf w}} = \Gamma {\bf z}$ holds, the user accepts $\hat{{\bf w}}$ as the result of $\Delta {\bf z}$. Otherwise, an error is declared and the verification process fails. This process has been summarized in Algorithm \ref{Alg:user_server}. In order to prove Theorem \ref{thm:interpol}, we propose the following lemma.

\begin{lemma}\textsf{INTERPOL} achieves the correctness and (information-theoretic) soundness properties defined in Section \ref{sec:statement}, with parameter ${\cal P}= q^{-c}$, ${\cal C} = O(c\sqrt{k})$ and $c_{\tilde{f}} = O(k)$.
\label{lemma:interpol}
\end{lemma}
The statements regarding the complexity and the correctness of $\textsf{INTERPOL}$ can be readily established. We will now prove the soundness property of the algorithm.

\subsection{Proof of Soundness}
Suppose that the server returns $\hat { {\bf w}}\neq { \bf w}$ where $\hat{{\bf w}}$ is an arbitrary (possibly randomized) function of ${\Delta}$ and ${\bf z}$. 
Our goal is to prove that $\hat{\bf w}$ only passes the verification check at the user with a negligible probability. More specifically, we have the following lemma. 
\begin{lemma}
Assume $\Delta\in \mathbb{F}_q^{s\times s}$ and ${\bf z}\in \mathbb{F}_q^s$ and let $\Lambda$  be a random matrix uniformly distributed over  $\mathbb{F}_q^{c\times s}$. Define $\Gamma = \Lambda \Delta$ and ${\bf w} = \Delta {\bf z}$. Let ${{\bf W}}$ be a random variable over $\mathbb{F}_q^s$, independent of $\Lambda$, with an arbitrary distribution $p_{{\bf W}}$. We have
\begin{eqnarray}
\mathbb{P}(\Lambda {\bf W} = \Gamma {\bf z}\; , {\bf W} \neq {\bf w}) \le q^{-c}.
\end{eqnarray}
\label{thm:security}
\end{lemma}
\begin{IEEEproof}
\begin{eqnarray}
\hspace{-10pt}&&\mathbb{P}(\Lambda{\bf W}= \Gamma {\bf z}, {\bf W} \neq {\bf w})\\
&=& \mathbb{P}(\Lambda ({\bf W} - {\bf w}) = {\bf 0} , {\bf W}\neq {\bf w}) \\
&=& \sum_{{\bf y}\in\mathbb{F}_q^s, {\bf y}\neq {\bf 0}}\mathbb{P}(\Lambda {\bf y} = {\bf 0})\mathbb{P}({\bf W} = {\bf w} + {\bf y})\\
&=& \sum_{{\bf y}\in\mathbb{F}_q^s, {\bf y}\neq {\bf 0}}q^{-c}\mathbb{P}({\bf W} = {\bf w} + {\bf y}) \le q^{-c}
\end{eqnarray}
where the last line follows from the fact that ${\Lambda {\bf y}} \sim \mbox{unif}({\mathbb F}_q^s)$ for any ${\bf y}\in \mathbb{F}_q^s$, ${\bf y}\neq {\bf 0}$. Furthermore, we can omit the conditioning in the third line due to the fact that $\Lambda $ is independent of ${\bf W}$. 
\end{IEEEproof}

 \begin{algorithm}
\caption[caption]{\textsf{INTERPOL}:  User-Server Setup}
\begin{algorithmic}[1]
\Statex {{\bf Input:} Vector $[a_0,\dots,a_{k-1}]$ and input vector ${\bf x}$ }
\Statex {\bf Output: }{Verification  and decoding vectors ${\bf ver},{\bf dec}$.}
\Statex{}
\Statex {\bf Initialization:}
 \State Define $\Delta = [a_{i,j}]_{s\times s}$ where $a_{i,j} = a_{is + j}$. 
\State Generate random matrix $\Lambda \in {\mathbb F}_q^{c\times s}$ and compute $\Gamma = \Lambda \Delta$. 
\State Reveal $\Delta$ to the server.
\Statex{}
\For {$i\in\{1,2,\dots ,size({\bf x})\}$}
\State Reveal $x = x_i$ to the server. 
\State Ask server to compute ${\bf w} = \Delta\begin{bmatrix} 1& x&\dots&
x^{s-1}\end{bmatrix}$.
\State Receive $\hat{\bf w}$ from server.
 \State $ver_i = \mathbf{1}\left(\Delta\hat{\bf w} = \Gamma \begin{bmatrix} 1& x&\dots&
x^{s-1}\end{bmatrix}\right)$.
\State $dec_i = \begin{bmatrix} 1& x^s&\dots&
x^{s(s-1)}\end{bmatrix}\hat{\bf w}$.
\EndFor
\State Return ${\bf ver},{\bf dec}$. 
\end{algorithmic}
\label{Alg:user_server}
\end{algorithm}


\subsection{Adaptivity of \textsf{INTERPOL}}
\label{sec:adaptivity}
So far we assumed that the user does not disclose the information about whether or not the results returned by the server are accepted. In certain circumstances it may be necessary to reveal this information publicly, for instance to nullify a contract between the user and the server. The question that we intend to address in this section is whether an adversarial server can exploit this information in order to increase his likelihood of bypassing the verification test in the subsequent rounds of computation. The short answer to this question is ``yes, but only marginally".\\
To give a formal answer, consider a setup where $m$ consecutive queries are made to a server. After each query, the user reveals to him whether his response has been accepted or rejected. We then ask what is the probability that the adversarial server can pass the verification test with a false result at least once in the $m$ rounds of interaction. As a benchmark, also consider a secondary server who receives no feedback about the verification process and who returns $m$ random outputs in response to the $m$ queries. The probability that this server can bypass the verification test at least once in $m$ rounds is given by $1 - (1-\frac{1}{q^c})^m$. We will show that our primary server who relies on the previous verification results can at most increase this probability to $\frac{m}{q^c}$.  
 \begin{theorem}
Suppose that we rely on \textsf{INTERPOL} to delegate the computation of a polynomial $f(x)$ of degree $k-1$ to a server for $m$ different inputs $x\in \{x_1,\dots,x_m\}$. The $m$ queries are made sequentially, and the output of the verification function is revealed publicly after each round. We have 
\begin{eqnarray*}
P_L = \mathbb{P}(\exists \ell\in[m] \mbox{ s.t. } ver_\ell = {1}, dec_\ell \neq f(x_\ell)) \le \frac{m}{q^c}.
\label{eqn:batch_L}
\end{eqnarray*}
\end{theorem}
The proof of this theorem follows immediately from the next Lemma, which we will prove in Appendix A.
\begin{lemma}
Assume $\Delta \in \mathbb{F}_q^{s\times s}$ and ${\bf z}_1,\dots,{\bf z}_m \in \mathbb{F}_q^s$ and let $\Lambda$ be a random matrix uniformly distributed over $\mathbb{F}_q^{c\times s}$. Define $\Gamma = \Lambda \Delta$ and ${\bf w}_\ell = \Delta {\bf z}_\ell$, $\ell\in[m]$. Let ${\bf W}_\ell$, $\ell\in [m]$ be $m$ random variables satisfying the Markov chain
\begin{eqnarray*}
\Lambda \longleftrightarrow (V_1,\dots,V_{\ell-1}) \longleftrightarrow {\bf W}_\ell
\end{eqnarray*}
where $V_\ell = \mathbf{1}( \Lambda {\bf W}_\ell =\Gamma{\bf z}_\ell)$. We have
\begin{eqnarray*}
\mathbb{P}(\exists \ell\in[m] \; s.t. \;  V_\ell = 1, {\bf W}_\ell \neq {\bf w}_\ell) \le \frac{m}{q^c}.
\end{eqnarray*}
\label{lemma:adaptivity}
\end{lemma}

\section{Description of multi-party \textsf{INTERPOL} and Proof of Theorem \ref{thm:interpol_multi}}
\label{sec:interpol_multi}
Similar to the user-server setup, we define $\Delta$ as the $s\times s$ matrix where $a_{i,j}= a_{i\times s + j}$. Furthermore, we define $\Delta_j$ as the submatrix of $\Delta$ which consists of rows $\{\frac{s(j-1)}{n},\dots, \frac{sj}{n}-1\}$. The algorithm works as follows.\\ 
Each node $\ell$ generates a set of $n-1$ random matrices $\Lambda_{\ell,j}\in{\mathbb {F}_q^{c\times \frac{s}{n}}}$ and performs a one-time computation of $\Gamma_{\ell,j} = \Lambda_{\ell,j} \Delta_j\; ,\; j\in [n]\backslash\{\ell\}$. At each round of the algorithm, each node $j$ will be in charge of computing ${\bf w}_j = \Delta_j \begin{bmatrix}
1&x&\dots&x^{s-1}
\end{bmatrix}$ which can be done in $O(\frac{s^2}{n})$. The result of these computations are broadcast to the network. Once node $\ell$ receives $\hat{\bf w}_{[n]\backslash\{\ell\}}$, he verifies the correctness of each $\hat{\bf w}_j$ by checking whether $\Gamma_{\ell,j}\begin{bmatrix}
1&x&\dots&x^{s-1}
\end{bmatrix} = \Lambda_{\ell,j}\hat{\bf w}_j$. Each of these verifications can be done in $O(\frac{cs}{n})$. Finally, assuming that all the verifications pass, node $\ell$ recovers $f(x) = \begin{bmatrix}
1 & x^s &\dots & x^{s(s-1)}
\end{bmatrix}\begin{bmatrix}
\hat{\bf w}_1^T\dots \hat{\bf w}_n^T
\end{bmatrix}^T$. The overall complexity of this algorithm is $O(\frac{s^2}{n} + csn + s)$ per node where the first term is the complexity of computing $\Delta_\ell \begin{bmatrix}1 & x&\dots & x^{s-1}\end{bmatrix}$ (as server), the second term is the complexity of performing the $n-1$ verifications and the last term is the complexity of recovering $f(x)$. Compared to a naive algorithm where each node individually computes $f(x)$, this algorithm is $\Theta(n)$ times faster, assuming that the number of nodes in the network is substantially smaller than $s$. This procedure has been summarized in Algorithm \ref{Alg:multi_party}. The performance of this algorithm is characterized in the next lemma which can be proven similarly to Lemma \ref{lemma:interpol}. Theorem \ref{thm:interpol_multi} follows immediately.
\begin{lemma}Multi-party \textsf{INTERPOL} achieves the correctness and (information-theoretic) soundness properties defined in Section \ref{sec:def_multi}, with parameter ${\cal P}= q^{-c}$ and ${\cal C} = O(\frac{k}{n} + cn\sqrt{k})$.
\end{lemma}

 \begin{algorithm}
\caption[caption]{\textsf{INTERPOL}: Multi-party Setup (algorithm ran by node $\ell \in [n]$).}
\begin{algorithmic}[1]
\Statex {{\bf Input:} Vector $[a_0,\dots,a_{k-1}]$ and input vector ${\bf x}$  }
 \Statex {\bf Output: }{Verification matrix and decoding vector ${\bf VER},{\bf dec}$}
 \Statex{}
 \Statex{\bf Initialization:}
 \State Define $\Delta = [a_{i,j}]_{s\times s}$ where $a_{i,j} = a_{is + j}$.
 \State {Define $\Delta_j$ as submatrice of $\Delta$ consisting of rows $\frac{s(j-1)}{n},\dots,\frac{sj}{n}-1$ for $j\in [n]$.}
 \State Generate $n-1$ random matrices $\Lambda_{\ell,j}\in{\mathbb F}_q^{c\times \frac{s}{n}}$ , $j\in [n]\backslash\{\ell\}$.
 \State{Compute $\Gamma_{\ell,j} = \Lambda_j \Delta_{\ell,j}$, $j\in [n]\backslash\{\ell\}$.}
\For { $i\in \{1,2,\dots, size({\bf x})\}$}
\State Let $x = x_i$.
 \Statex{\bf As Server:}
 \State Compute ${\bf w}_\ell = {\Delta}_\ell \begin{bmatrix} 1 & x & \dots & x^{s-1}\end{bmatrix}$.
 \State Broadcast ${\bf w}_\ell$ to every other node.
 \Statex{\bf As User:}
 \For {$j\in [n]\backslash\{\ell\}$}
 \State Receive $\hat{\bf w}_j$ from server $j$.
\State $\mbox{VER}_{i,j} = \mathbf{1}\left(\Lambda_{\ell,j}\hat{\bf w}_j = \Gamma_{\ell,j} \begin{bmatrix} 1 & x_i & \dots & x_i^{s-1}\end{bmatrix} \right)$.
 \EndFor
  \State $\mbox{dec}_i = \begin{bmatrix} 1 & x_i^s & \dots & x_i^{s(s-1)}\end{bmatrix} \begin{bmatrix} \hat{\bf w}_1^T & \dots& \hat{\bf w}_n^T\end{bmatrix}^T $.
\EndFor
\State Return ${{\bf VER}},{\bf dec}$. 
\end{algorithmic}
\label{Alg:multi_party}
\end{algorithm}
Given that multi-party \textsf{INTERPOL} is a straightforward generalization of user-server \textsf{INTERPOL}, it is not hard to see that it also preserves the adaptivity feature. We omit this analysis for conciseness.
\subsection{Enabling Error Correction in the Multi-party Setting}
As suggested in Section \ref{sec:main}, in the multi-party setting the network can dynamically remove the malicious nodes and repeat the computation that was assigned to them. An alternative approach to redoing the computation is to use erasure coding. Specifically, in Algorithm \ref{Alg:multi_party}, we can divide the matrix $\Delta$ into $k$ horizontal sub-matrices, and apply an $(n,k)$ Reed-Solomon code to these submatrices to obtain $\{\tilde{\Delta}_1,\tilde{\Delta}_2,\dots,\tilde{\Delta}_n\}$. We can then assign the task of computing $\tilde{\Delta}_\ell \begin{bmatrix}1 &x &\dots & x^{s-1}\end{bmatrix}$ to node $\ell$. This approach increases the overall complexity of each node by a factor of $\frac{n}{k}$. However, in the presence of up to $n-k$ malicious nodes, each node can recover all $\tilde{\Delta}_\ell \begin{bmatrix}1 &x &\dots & x^{s-1}\end{bmatrix}$ efficiently, thanks to the fast-decodability of RS codes. To appreciate the role of \textsf{INTERPOL}, note that in the absence of a verification mechanism, the same algorithm would only tolerate $\lceil\frac{n-k+1}{2}\rceil - 1$ errors (albeit in a deterministic fashion). With the help of \textsf{INTERPOL}, we can locate the errors, thereby increase the tolerance against malicious servers by a factor of 2.  

\section{Generalization to Multivariate Polynomials}
Consider the following $m$-variate polynomial which is of degree $n-1$ in each variable.
\begin{align}
    f(x_1,\dots,x_m) = \sum_{d_1\in[0:n-1],\dots,d_m\in[0:n-1]}a_{d_1,\dots,d_m}\prod_{i=1}^{m}x_i^{d_i}.
\end{align}
Evaluating this polynomial can be done in time linear in the number of non-zero terms, which in the worst case scenario is $O(n^m)$. Suppose for simplicity that $m$ is even. We will propose an extension of INTERPOL that reduces the complexity of the user to $O(n^{\frac{m}{2}})$. We start by partitioning the set of variables into two sets $\{x_1,\dots,x_{m/2}\}$ and $\{x_{m/2+1},\dots, x_m\}$. We can rewrite $f(x_1,\dots,x_m)$ as  
\begin{eqnarray} 
f(x_1,\dots,x_m) = 
{\bf x}_0^T \Delta {\bf x}_1
\end{eqnarray}
where each ${\bf x}_0$ and ${\bf x}_1$ is a vertical vector of length $(n^{\frac{m}{2}})$ and  
\begin{eqnarray}
{\bf x}_\ell[i] = \prod_{j=1}^{m/2}x_{j+\ell m/2}^{b_{i,j}}, \;\; \forall i\in [0:n^{\frac{m}{2}}-1], \ell\in[0:1],
\end{eqnarray}
where $[b_{i,1},\dots,b_{i,m/2}]$ is the $n$-ary expansion of $i$. Furthermore, $\Delta$ is an $n^{m/2}\times n^{m/2}$ matrix whose elements are \begin{eqnarray}
\Delta_{ij} = a_{{\bf d}_{i,j}}
\end{eqnarray}
where 
${\bf d}_{i,j} := [b_{i,1},\dots,b_{i,m/2},b_{j,1},\dots,b_{j,m/2}].$

The user can now perform a one-time heavy computation of $\Gamma = \Lambda \Delta$ where $\Lambda$ is a secret $c\times n^{m/2}$ matrix uniformly generated at random over the field. The user delegates the computation of ${\bf w} = \Delta {\bf x}_1$ to the server. After receiving $\hat{\bf w}$, he can check whether $\Lambda \hat{\bf w} = \Gamma {\bf x}_1$. Once the verification process has passed, the user will compute $f(x_1,\dots,x_m) = {\bf x}_0^T {\bf w}$. The complexity of the user will be $O(n^{m/2})$ while the complexity of the server remains $O(n^m)$.  
\section*{Acknowledgement}
We thank support from the Distributed Technologies  Research Foundation. This work is also supported by Defense Advanced Research Projects Agency (DARPA) under Contract No. HR001117C0053. The views, opinions, and/or findings expressed are those of the author(s) and should not be interpreted as representing the official views or policies of the Department of Defense or the U.S. Government. This work is also supported by ARO award W911NF-18-1-0400,  NSF Grants CCF-1763673 and CCF-1703575, and the Swiss National Science Foundation Grant 178309. 

\bibliographystyle{IEEEtran}
\bibliography{IEEEfull,main}

\begin{thebibliography}{10}
\providecommand{\url}[1]{#1}
\csname url@samestyle\endcsname
\providecommand{\newblock}{\relax}
\providecommand{\bibinfo}[2]{#2}
\providecommand{\BIBentrySTDinterwordspacing}{\spaceskip=0pt\relax}
\providecommand{\BIBentryALTinterwordstretchfactor}{4}
\providecommand{\BIBentryALTinterwordspacing}{\spaceskip=\fontdimen2\font plus
\BIBentryALTinterwordstretchfactor\fontdimen3\font minus
  \fontdimen4\font\relax}
\providecommand{\BIBforeignlanguage}[2]{{%
\expandafter\ifx\csname l@#1\endcsname\relax
\typeout{** WARNING: IEEEtran.bst: No hyphenation pattern has been}%
\typeout{** loaded for the language `#1'. Using the pattern for}%
\typeout{** the default language instead.}%
\else
\language=\csname l@#1\endcsname
\fi
#2}}
\providecommand{\BIBdecl}{\relax}
\BIBdecl

\bibitem{gennaro2010non}
R.~Gennaro, C.~Gentry, and B.~Parno, ``Non-interactive verifiable computing:
  Outsourcing computation to untrusted workers,'' in \emph{Annual Cryptology
  Conference}.\hskip 1em plus 0.5em minus 0.4em\relax Springer, 2010, pp.
  465--482.

\bibitem{benabbas2011verifiable}
S.~Benabbas, R.~Gennaro, and Y.~Vahlis, ``Verifiable delegation of computation
  over large datasets,'' in \emph{Annual Cryptology Conference}.\hskip 1em plus
  0.5em minus 0.4em\relax Springer, 2011, pp. 111--131.

\bibitem{babai1985trading}
L.~Babai, ``Trading group theory for randomness,'' in \emph{Proceedings of the
  seventeenth annual ACM symposium on Theory of computing}.\hskip 1em plus
  0.5em minus 0.4em\relax ACM, 1985, pp. 421--429.

\bibitem{kilian1992note}
J.~Kilian, ``A note on efficient zero-knowledge proofs and arguments,'' in
  \emph{Proceedings of the twenty-fourth annual ACM symposium on Theory of
  computing}.\hskip 1em plus 0.5em minus 0.4em\relax ACM, 1992, pp. 723--732.

\bibitem{goldwasser2008delegating}
S.~Goldwasser, Y.~T. Kalai, and G.~N. Rothblum, ``Delegating computation:
  interactive proofs for muggles,'' in \emph{Proceedings of the fortieth annual
  ACM symposium on Theory of computing}.\hskip 1em plus 0.5em minus 0.4em\relax
  ACM, 2008, pp. 113--122.

\bibitem{parno2013pinocchio}
B.~Parno, J.~Howell, C.~Gentry, and M.~Raykova, ``Pinocchio: Nearly practical
  verifiable computation,'' in \emph{2013 IEEE Symposium on Security and
  Privacy}.\hskip 1em plus 0.5em minus 0.4em\relax IEEE, 2013, pp. 238--252.

\bibitem{gennaro2013quadratic}
R.~Gennaro, C.~Gentry, B.~Parno, and M.~Raykova, ``Quadratic span programs and
  succinct {NIZK}s without {PCP}s,'' in \emph{Annual International Conference
  on the Theory and Applications of Cryptographic Techniques}.\hskip 1em plus
  0.5em minus 0.4em\relax Springer, 2013, pp. 626--645.

\bibitem{elkhiyaoui2016efficient}
K.~Elkhiyaoui, M.~{\"O}nen, M.~Azraoui, and R.~Molva, ``Efficient techniques
  for publicly verifiable delegation of computation,'' in \emph{Proceedings of
  the 11th ACM on Asia Conference on Computer and Communications
  Security}.\hskip 1em plus 0.5em minus 0.4em\relax ACM, 2016, pp. 119--128.

\bibitem{fiore2012publicly}
D.~Fiore and R.~Gennaro, ``Publicly verifiable delegation of large polynomials
  and matrix computations, with applications,'' in \emph{Proceedings of the
  2012 ACM conference on Computer and communications security}.\hskip 1em plus
  0.5em minus 0.4em\relax ACM, 2012, pp. 501--512.

\bibitem{zhang2017new}
X.~Zhang, T.~Jiang, K.-C. Li, A.~Castiglione, and X.~Chen, ``New publicly
  verifiable computation for batch matrix multiplication,'' \emph{Information
  Sciences}, 2017.

\bibitem{polyshard}
S.~Li, M.~Yu, S.~Avestimehr, S.~Kannan, and P.~Viswanath, ``Polyshard: Coded
  sharding achieves linearly scaling efficiency and security simultaneously,''
  \emph{arXiv preprint arXiv:1809.10361}, 2018.

\bibitem{yao1982protocols}
A.~C. Yao, ``Protocols for secure computations,'' in \emph{Foundations of
  Computer Science, 1982. SFCS'08. 23rd Annual Symposium on}.\hskip 1em plus
  0.5em minus 0.4em\relax IEEE, 1982, pp. 160--164.

\bibitem{lindell2009proof}
Y.~Lindell and B.~Pinkas, ``A proof of security of {Y}ao’s protocol for
  two-party computation,'' \emph{Journal of Cryptology}, vol.~22, no.~2, pp.
  161--188, 2009.

\bibitem{zhang2014efficient}
Y.~Zhang and M.~Blanton, ``Efficient secure and verifiable outsourcing of
  matrix multiplications,'' in \emph{International Conference on Information
  Security}.\hskip 1em plus 0.5em minus 0.4em\relax Springer, 2014, pp.
  158--178.

\bibitem{chen2014new}
X.~Chen, J.~Li, J.~Ma, Q.~Tang, and W.~Lou, ``New algorithms for secure
  outsourcing of modular exponentiations,'' \emph{IEEE Transactions on Parallel
  and Distributed Systems}, vol.~25, no.~9, pp. 2386--2396, 2014.

\bibitem{backes2013verifiable}
M.~Backes, D.~Fiore, and R.~M. Reischuk, ``Verifiable delegation of computation
  on outsourced data,'' in \emph{Proceedings of the 2013 ACM SIGSAC conference
  on Computer \& communications security}.\hskip 1em plus 0.5em minus
  0.4em\relax ACM, 2013, pp. 863--874.

\end{thebibliography}

\section*{Appendix A: Proof of Lemma \ref{lemma:adaptivity}}
Without loss of generality, suppose the adversarial server chooses ${\bf W}_\ell = {\bf w}_\ell + {\bf Y}_\ell$ where ${\bf Y}_\ell$ is an arbitrary random variable defined over $F_q^{s}\backslash \{{\bf 0}\}$ which satisfies the Markov chain
\begin{eqnarray}
\Lambda \longleftrightarrow (V_1,\dots,V_{\ell-1}) \longleftrightarrow {\bf Y}_\ell
\label{eqn:second_markov}
\end{eqnarray}
where $V_\ell = \mathbf{1}( \Lambda {\bf Y}_\ell ={\bf 0})$. We can compute
\begin{eqnarray*}
p_m &:=&\mathbb{P}(V_1 = 0,\dots, V_m = 0)\\
&=&\mathbb{P}(\Lambda {\bf Y}_m \neq {\bf 0},\dots, \Lambda {\bf Y}_1 \neq {\bf 0}) \\
&=& \mathbb{P}(\Lambda {\bf Y}_m \neq {\bf 0}|\Lambda {\bf Y}_{m-1}\neq {\bf 0},\dots, \Lambda {\bf Y}_1 \neq {\bf 0})\\
&\times&\mathbb{P}(\Lambda {\bf Y}_{m-1} \neq {\bf 0},\dots, \Lambda {\bf Y}_1 \neq {\bf 0}).
\end{eqnarray*}
Let us bound the first term.
\begin{align*}
t_m &:= \mathbb{P}(\Lambda {\bf Y}_m \neq {\bf 0}|\Lambda {\bf Y}_{m-1}\neq {\bf 0},\dots, \Lambda {\bf Y}_1 \neq {\bf 0}) \\ 
&= \sum_{{\bf y}_m \neq {\bf 0}}\mathbb{P}(\Lambda {\bf Y}_m \neq {\bf 0}| {\bf Y}_m = {\bf y}_m, \Lambda {\bf Y}_{m-1}\neq {\bf 0}\dots,\\
& \Lambda {\bf Y}_1 \neq {\bf 0})
\times\mathbb{P}({\bf Y}_m = {\bf y}_m | \Lambda {\bf Y}_{m-1}\neq {\bf 0},\dots, \Lambda {\bf Y}_1 \neq {\bf 0})\\
&= \sum_{{\bf y}_m \neq {\bf 0}}\mathbb{P}(\Lambda {\bf y}_m \neq {\bf 0}| \Lambda {\bf Y}_{m-1}\neq {\bf 0},\dots, \Lambda {\bf Y}_1 \neq {\bf 0})\\
&\times\mathbb{P}({\bf Y}_m = {\bf y}_m| \Lambda {\bf Y}_{m-1}\neq {\bf 0},\dots, \Lambda {\bf Y}_1 \neq {\bf 0} )
\end{align*}
where the last equality follows from the Markov chain property \eqref{eqn:second_markov}. We proceed by proposing a bound on the term $\mathbb{P}(\Lambda {\bf y}_m \neq {\bf 0}| \Lambda {\bf Y}_{m-1}\neq {\bf 0},\dots, \Lambda {\bf Y}_1 \neq {\bf 0})$ that holds for any ${\bf y}_m \neq {\bf 0}$.  
\begin{eqnarray*}
r_m &:=& \mathbb{P}(\Lambda {\bf y}_m \neq {\bf 0}| \Lambda {\bf Y}_{m-1}\neq {\bf 0},\dots, \Lambda {\bf Y}_1 \neq {\bf 0})\\
&=& 1- \mathbb{P}(\Lambda {\bf y}_m = {\bf 0}| \Lambda {\bf Y}_{m-1}\neq {\bf 0},\dots, \Lambda {\bf Y}_1 \neq {\bf 0})\\
&=& 1-\frac{\mathbb{P}(\Lambda {\bf y}_m= {\bf 0}, \Lambda {\bf Y}_{m-1}\neq {\bf 0},\dots, \Lambda {\bf Y}_1 \neq {\bf 0})}{\mathbb{P}(\Lambda {\bf Y}_{m-1}\neq {\bf 0},\dots, \Lambda {\bf Y}_1 \neq {\bf 0})}\\
&\ge & 1- \frac{\mathbb{P}(\Lambda {\bf y}_m = {\bf 0})}{\mathbb{P}( \Lambda {\bf Y}_{m-1}\neq {\bf 0},\dots, \Lambda {\bf Y}_1 \neq {\bf 0})} \\
&=& 1 - \frac{1}{q^c \mathbb{P}(\Lambda {\bf Y}_{m-1}\neq {\bf 0},\dots, \Lambda {\bf Y}_1 \neq {\bf 0})}.
\end{eqnarray*}
We can now continue with lower-bounding $t_m$ as
\begin{eqnarray*}
t_m &\ge& \sum_{{\bf y}_m \neq {\bf 0}} \left(1 - \frac{1}{q^c \mathbb{P}(\Lambda {\bf Y}_{m-1}\neq {\bf 0},\dots, \Lambda {\bf Y}_1 \neq {\bf 0})}\right)\\
&\times&\mathbb{P}({\bf Y}_m = {\bf y}_m| \Lambda {\bf Y}_{m-1}\neq {\bf 0},\dots, \Lambda {\bf Y}_1 \neq {\bf 0} )\\
&=&1 - \frac{1}{q^c \mathbb{P}(\Lambda {\bf Y}_{m-1}\neq {\bf 0},\dots, \Lambda {\bf Y}_1 \neq {\bf 0})}.
\end{eqnarray*}
Finally, we can lower-bound $p_m$ as 
\begin{eqnarray*}
p_m &\ge& (1 - \frac{1}{q^c \mathbb{P}(\Lambda {\bf Y}_{m-1}\neq {\bf 0},\dots, \Lambda {\bf Y}_1 \neq {\bf 0})})\\
&\times& \mathbb{P}(\Lambda {\bf Y}_{m-1}\neq {\bf 0},\dots, \Lambda {\bf Y}_1 \neq {\bf 0})\\
&=& p_{m-1} - \frac{1}{q^c}.
\end{eqnarray*}
This recursion and the fact that $p_1 \ge 1- \frac{1}{q^c}$ result in our desired bound $p_m \ge 1 - \frac{m}{q^c}$.
\end{document}